 \pdfoutput=1

\documentclass[prl, aps, twocolumn,longbibliography,notitlepage]{revtex4}
\usepackage{graphicx} 
\usepackage[usenames]{color}
\usepackage{amsmath,amssymb}
\usepackage{gensymb}
\usepackage{natbib}
\usepackage{tabularx}
\usepackage{xcolor}
\usepackage{upgreek}
\usepackage{soul}
\usepackage{hyperref}
\usepackage{subfigure}
\usepackage{changes}

\usepackage{bbm}

\begin{document}

\title{{Fiber buckling in confined viscous flows: an absolute instability described by the Ginzburg-Landau equation}}

\author{Jean Cappello$^{1}$, Olivia du Roure$^{1}$, Fran\c cois Gallaire$^2$, Camille Duprat$^{3}$,  and Anke Lindner$^{1}$}
\affiliation {$^{1}$ PMMH, ESPCI, Paris, France\\
$^{2}$ LMFI, EPFL, Lausanne, Switzerland\\
$^{3}$ LadHyX, \'Ecole Polytechnique, Palaiseau, France}
%___________________________________________________________________________
\begin{abstract}
We explore the dynamics of a flexible fiber transported by a viscous flow in a Hele-Shaw cell of height comparable to the fiber height. We show that long fibers aligned with the flow experience a buckling instability. Competition between viscous and elastic forces leads to the deformation of the fiber into a wavy shape convolved by a Bell-shaped envelope. We characterize the wavelength and phase velocity of the deformation as well as the growth and spreading of the envelope. Our study of the spatio-temporal evolution of the deformation reveals a linear and absolute instability arising from a local mechanism well described by the Ginzburg-Landau equation.

\end{abstract}
%
%\pacs{83.80.Hj,47.57.Gc,47.57.Qk,82.70.Kj}
%
\date{\today}
\maketitle

The dynamics of a flexible fiber transported in a flow results from a coupling between deformation and transport and exhibits rich behaviours \cite{duRoure2019} including bending \cite{Cappello2019, Marchetti2018, Li2013}, buckling \cite{Guglielmini2012, Li2013, Young2007, Wandersman2010} or coiling \cite{Chakrabarti2020}. Such dynamics have been studied in different types of flows including simple shear flows \cite{Liu2018, Forgacs1959, Harasim2013, Nguyen2014, Becker2001}, vortex arrays \cite{Young2007, Wandersman2010, Quennouz2015, Manikantan2013}, Poiseuille flows \cite{Steinhauser2012}, extensional flows \cite{Guglielmini2012, Manikantan2015, Liu2020, Chakrabarti2020} or sedimentation in quiescent fluids \cite{Xu1994, Li2013, Marchetti2018}. The transport dynamics of flexible fibers in confined geometries has only recently been explored \cite{DAngelo2009_1, dAngelo2009, Cappello2019}, and shows even richer dynamics, as a consequence of the interaction between the fiber of evolving shape, and the confining walls.

 \begin{figure}
\centering
\includegraphics[width=0.9 \columnwidth]{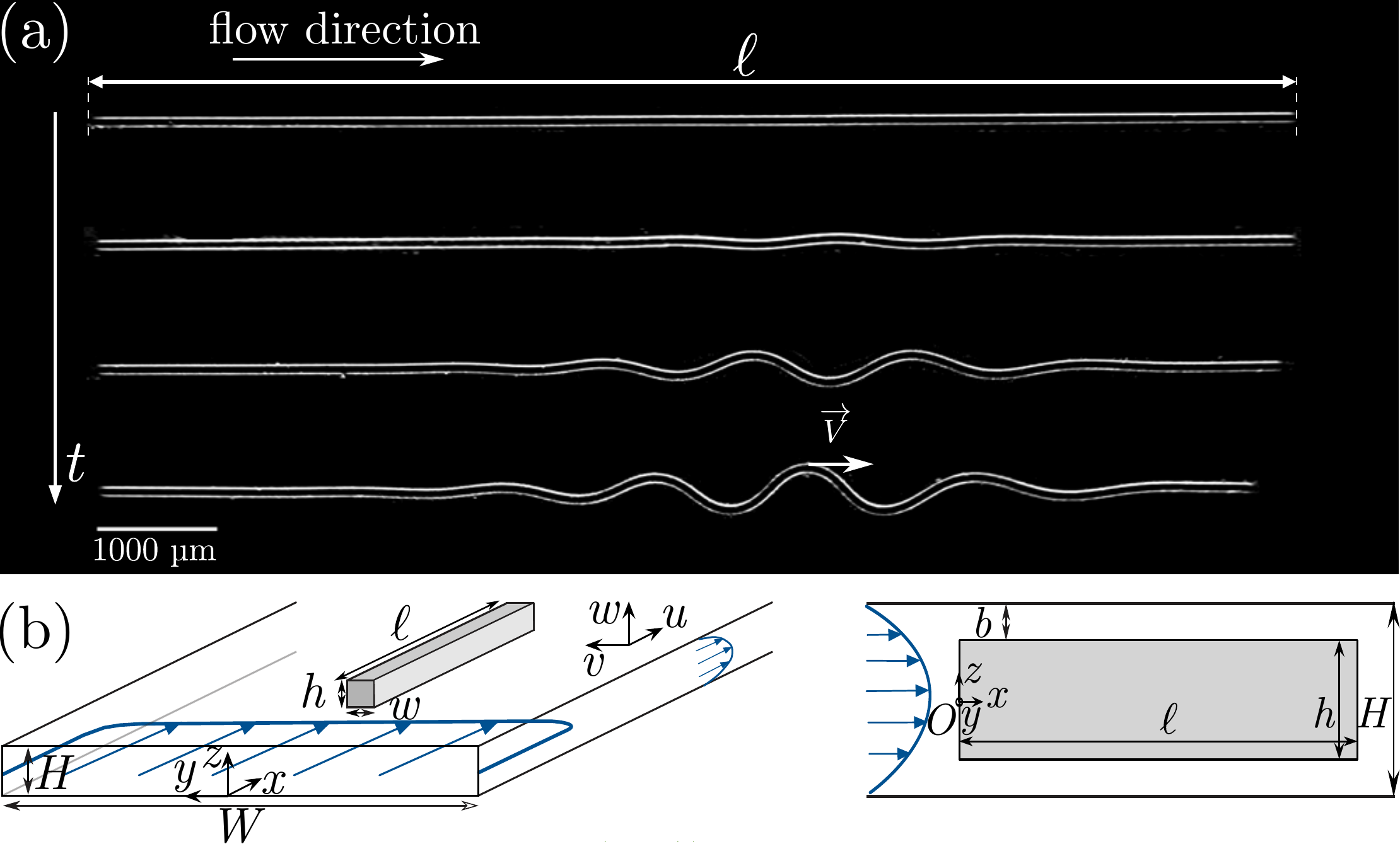}
\caption{(a) Chronophotographies of a parallel fiber transported by the external flow in the moving reference frame of the fiber. Time increases from top to bottom (time interval is 4 sec). The phase velocity is denoted $\vec{V}$ and the horizontal arrow illustrates the direction of the perturbation propagation. The flow velocity, from left to right, is $u_0 = 0.65$ mm$\cdot$s$^{-1}$. The fiber has a length $\ell = 10130 \pm 15$ $\upmu$m, width $w=75 \pm 3$ $\upmu$m, height $h=73\pm3$ $\upmu$m, Young's modulus $E= 27 \pm7$ kPa. The fluid viscosity is $\mu = 85 \pm 3$ mPa$\cdot$s. Channel height and width are respectively  $H=85\pm3$ $\upmu$m and $W=3500$ $\upmu$m. Scale bar is 1000 $\upmu$m. (b) Geometry of the fiber and the confining channel. The pressure-driven flow is sketched in blue.}
\label{fig:figure_1}
\end{figure}  

In this study, we focus on the dynamics of a flexible fiber transported by a pressure-driven flow in a Hele-Shaw cell (see Fig.~\ref{fig:figure_1}). The fiber is aligned with the flow and is confined by the channel top and bottom walls. Typical experimental observations can be seen in Fig. \ref{fig:figure_1}~(a) showing the shape of a flexible fiber during transport along the microchannel. The fiber, initially straight, deforms into a sinusoidal shape with a very well defined wavelength modulated by a bell-shaped envelope. The dynamics of the deformation is manifold: the amplitude of the deformation increases with time, the perturbation travels along the fiber from back to front and  the envelope broadens along the fiber while remaining centered at a fixed lengthwise position. 

Such dynamical shape changes can be triggered through various mechanisms amplifying initial perturbations, classified as either global or local instabilities.
Global instabilities span the entire domain (here the fiber) and couple the two end conditions. Hence the boundary conditions are essential in the selection of the deformation mode. This is typically the case for classical buckling instabilities where the perturbation ranges over the entire fibre and the perturbation wavelength is function of the fiber length \cite{duRoure2019}.

On the contrary, in the presence of a local amplifying mechanism the deformations result from an initial disturbance invading the unstable system via propagating fronts. If the perturbation grows but is simultaneously advected such that the disturbance eventually decays at any fixed point, it ultimately leaves the fiber and the instability is said to be convective. In contrast, as is the case here, if the perturbation grows at a fixed spatial position, it eventually invades the entire fiber leading to a self-sustained instability and is said to be absolute. %Note that a full invasion of the system by an absolute instability is sometimes also referred to as a global instability, but here it originates from the amplification of a local perturbation and results in an intrinsic, size-independent wavelength selection.

{Absolute} instabilities have been observed in open flow hydrodynamics, such as the flow past bluff bodies (B\'enard-von K\'arm\'an vortex street)  \cite{Triantafyllou1987}, parallel shear flows \cite{Gondret1999, Charru2012,  Huerre2000}, transition from dripping to jetting \cite{Utada2008, Gordillo2001, Ganan2006, Guillot2007} or the flow of viscous films down vertical fibers \cite{Duprat2007, Gallaire2017}. However, absolute instabilities {resulting} in the deformation of a flexible object in interaction with viscous flows have not been reported yet. In this letter we describe a one-dimensional example of such an instability developing during the transport of a flexible fiber in a viscous flow.

We first characterize the spatial and temporal evolution of the observed wave by studying the wavelength and the phase velocity of the perturbation as a function of the physical parameters of the system. Then, by the means of the linear Ginzburg-Landau model - the archetypal evolution model leading to  an absolute or a convective instability \cite{Huerre1990} - we describe the spatio-temporal dynamics of the envelope {and show} the linear and local nature of the absolute instability as well as a near zero group velocity.

\paragraph{Material and methods.}

 \begin{figure}[ht]
\centering
\includegraphics[width= \columnwidth]{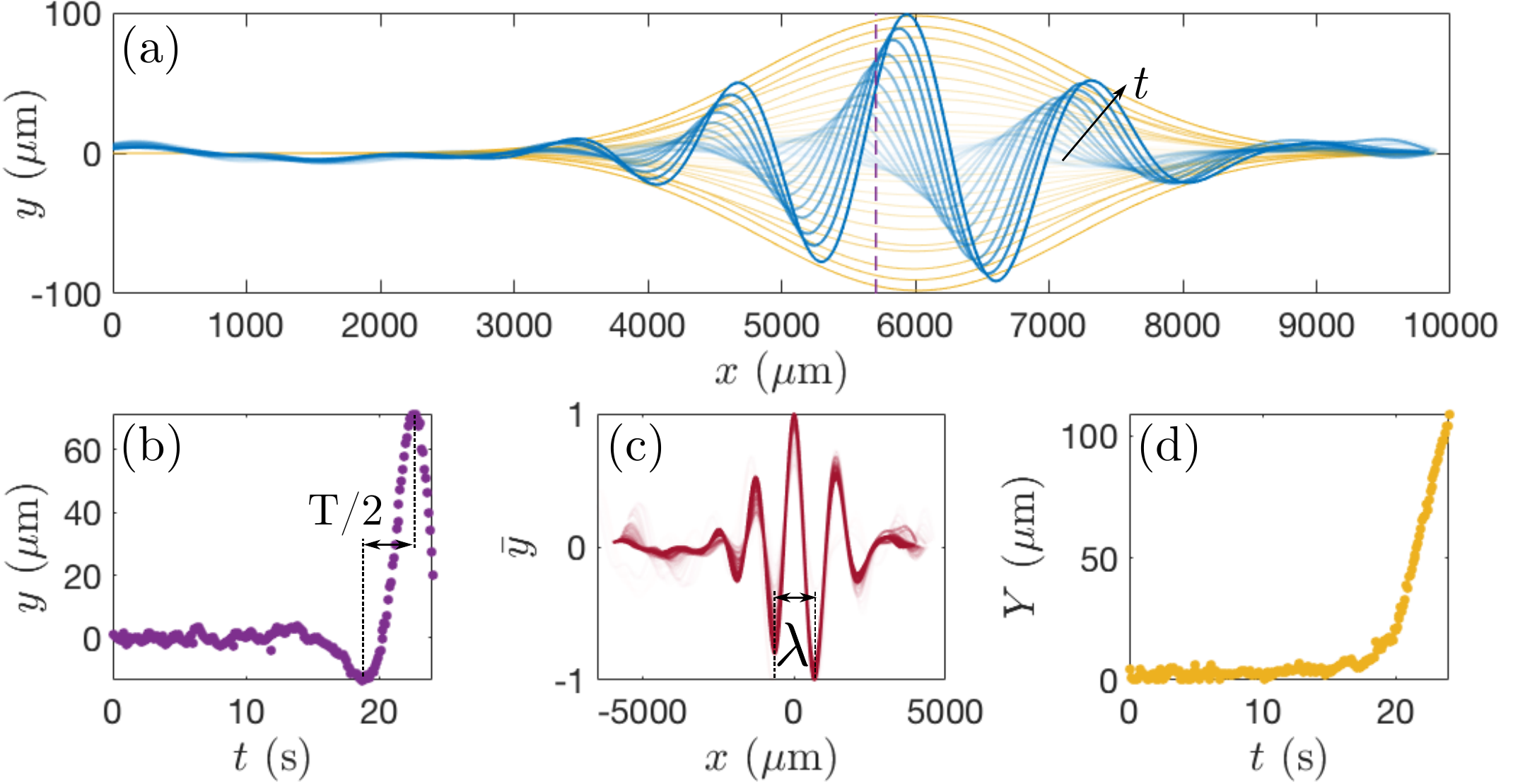}
\caption{Characterization of the fiber deformation. (a) Blue: Superposition of the shapes of the fiber shown in Fig.~\ref{fig:figure_1} in the reference frame of the fiber. Yellow: Superposition of the envelope of the deformation. Darker colors indicate later time points (time interval is 0.4 sec). (b) Temporal evolution of the deformation amplitude at a fixed $x$-location on the fiber (purple dashed line in (a)). The period of the oscillation, $T$, can be extracted from this curve. (c)  Superposition of shapes normalized by their maximum {($\bar{y} = y / {\rm{max}}(y)$)} and aligned on their maxima ($x=0$). The wavelength of the deformation, $\lambda$, is constant. (d) Growth of the envelope amplitude with time.}
\label{fig:figure_2}
\end{figure}

Fibers are fabricated in a microchannel of Hele-Shaw geometry (see Figure~\ref{fig:figure_1}~(b)), using the stop-flow microscope-based projection photo-lithography method \cite{Dendukuri2008, Dendukuri2007}. This method enables the fabrication, directly inside a microchannel, of polymeric PEG-based hydrogel particles of rectangular cross section, surrounded by a Newtonian solution of uncured PEGDA. It allows for an excellent control of the particle  geometry and its mechanical properties \cite{Duprat2014, Cappello2019, Cappello2020}.  Details can be found in references \cite{Dendukuri2008, Wexler2013, Berthet2016, Duprat2014, Nagel2018, Cappello2019, Bechert2019, Cappello2020}. An important feature of this {method} is the existence of a liquid gap between the fiber and the channel top and bottom walls of constant height $b = 6.0\pm1.6$ $\upmu$m. We use channels of fixed height $H=85\pm3$ $\upmu$m and different widths ($W=1000$ $\upmu$m - 3500 $\upmu$m). The resulting fiber height is $h=73\pm3$ $\upmu$m {leading to strong confinement by the top and bottom walls: $h/H=0.86$. Fiber width varies from $w=73$ $\upmu$m to $w=118$ $\upmu$m while the length varies from $\ell =9174 \pm 15$ $\upmu$m to $\ell =15000 \pm 15$ $\upmu$m and Young modulus varies from {$E=117\pm64$ kPa to $E=27 \pm7$ kPa}.}

Once a fiber is fabricated, flow is turned on with a syringe pump (Nemesys, Cetoni). Because of the Hele-Shaw geometry of the channel, the flow is plug-like in the ($Oxy$) plane, except in the vicinity of the lateral walls and of the fiber, and Poiseuille like in the ($Oxz$) plane (see Fig.~\ref{fig:figure_1}~(b)).
The fiber  is imaged using  an Hamamatsu Orca-flash 4.0 camera, at 10 fps. The fiber is kept in the camera's field of view by manually moving the microscope stage.

The fiber shapes are extracted from images by standard image processing (with  ImageJ  \cite{Schneider2012} and Matlab). 
The wavelength is measured by spatial Fourier transform while the envelope is obtained through an Hilbert transform.

\paragraph{Results}

The evolution of the shape of the fiber shown in Fig.~\ref{fig:figure_1}~(a) is summarized in Fig.~\ref{fig:figure_2}~(a). The fiber adopts a wavy shape (blue curves) whose amplitude decreases toward the edges as emphasized by the shape envelopes (plotted in yellow). Darker curves correspond to later time points and one can see that the deformation amplitude increases with time. In Fig.~\ref{fig:figure_2}~(a) the back ends have been aligned to highlight that the wave propagates along the fiber from back to front at the phase velocity $V = \lambda/T$ with $\lambda$ the wavelength and $T$ the wave period. The wave period is measured by following the deformation amplitude $y$ on a fixed abscissa as shown in Fig.~\ref{fig:figure_2}~(b). The wavelength is shown to be constant by superposing normalized shapes aligned on their maxima (Fig.~\ref{fig:figure_2}~(c)). As time increases, the deformation increases and spreads along the fiber (see Fig.~\ref{fig:figure_2}~(a) and (d) and Figs.~\ref{fig:GL}~(a) and (c)) until saturation (not shown). The evolution of the fiber velocity is given in Fig.~1 of the supp.  data. After a transient regime the fiber velocity reaches a plateau value that precedes the mechanical instability.

\begin{figure}
\centering
\includegraphics[width=0.9 \columnwidth]{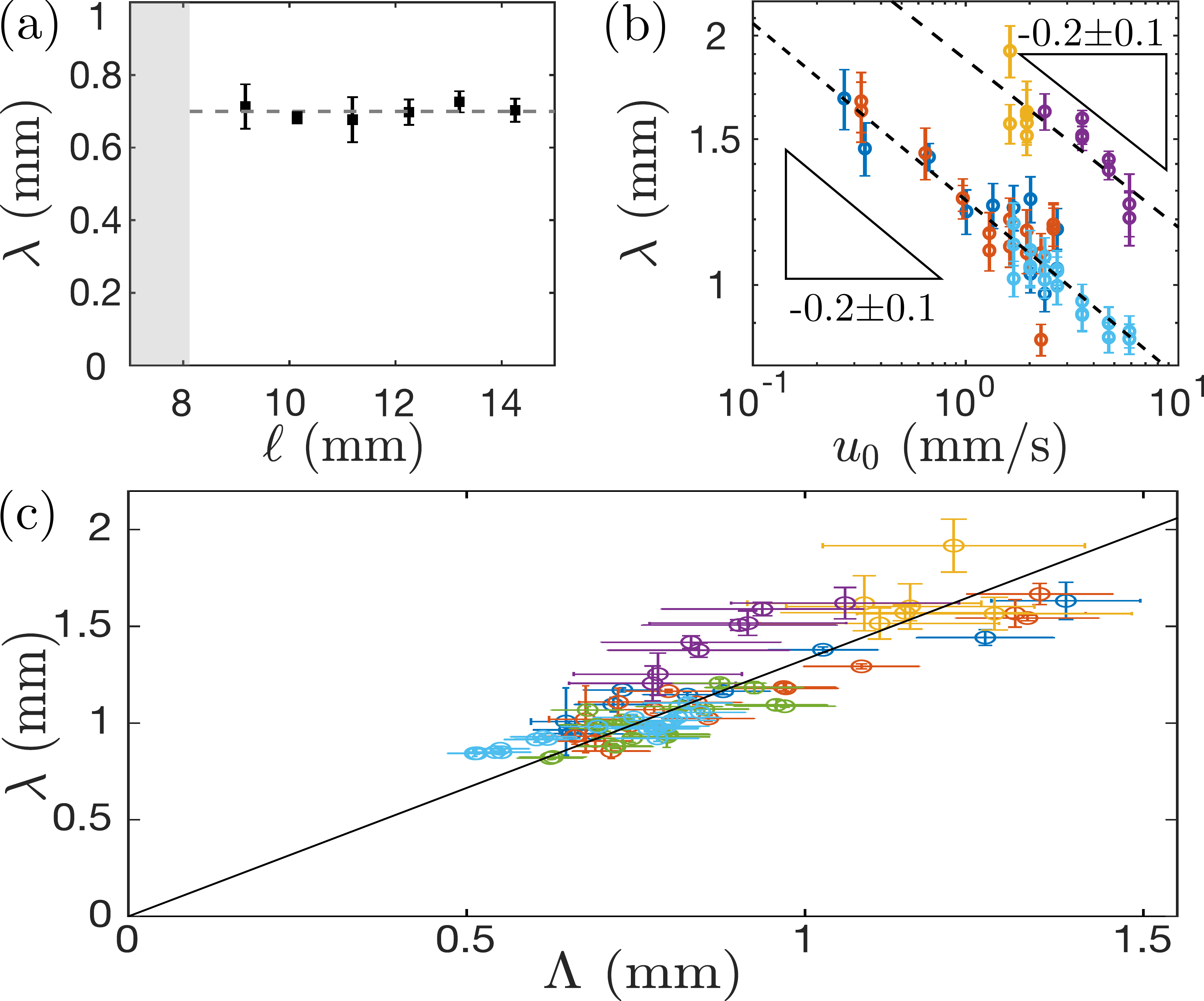}
\caption{Wavelength characterization. (a) Evolution of the wavelength $\lambda$ as a function of the fiber length $\ell$ (all other parameters are kept constant. Fiber: $E = 0.20\pm 0.11$ kPa, $w=65\pm3\upmu$m, $h=73 \pm3 \upmu$m. Channel: $W=3500$ $\upmu$m, $H=85$ $\upmu$m.). (b) Evolution of $\lambda$ as a function of $u_0$. Two sets of data are represented gathering 59 experiments with different Young moduli, fiber lengths and channel widths, other parameter being constant ($w = 75\pm9$ $\upmu$m, $h = 73\pm3$ $\upmu$m, $H=85 \pm 3$ $\upmu$m). Yellow and purple: $E = 117\pm64$ kPa, $\ell = 15.64\pm 0.47$mm. Red, dark blue and light blue: $E = 27\pm7$ kPa, $\ell= 10.187 \pm 0.025$ mm. Red: $W = 3640 \pm 10$ $\upmu$m; yellow and dark blue: $W = 3500 \pm 10$ $\upmu$m; purple and light blue: $W = 1000\pm10$ $\upmu$m. Dotted black lines correspond to fits by a power law. (c) Evolution of the wavelength as a function of the elasto-viscous length $\Lambda$. The data set is the same as in (b) (same color code) and has been complemented by the green points corresponding to variation of the width (from $73$ $\upmu$m to $118$ $\upmu$m), other parameters being constant ($E = 27 \pm 7$ kPa, $\ell= 10.23 \pm 0.08$ mm, $h = 73 \pm 3$ $\upmu$m {and $u_0 = 3.53$ mm/s}). In total 72 measurements are shown in the figure. The dotted line corresponds to a linear fit of equation $\lambda$ = 1.33 $\Lambda$.}
\label{fig:wavelength_evo}
\end{figure}

 To study the instability we first focus on the spatial and temporal evolution of the signal by characterizing the wavelength and the phase velocity of the wave. We characterize the dependence of the wavelength on the different parameters of the problem in Fig.~\ref{fig:wavelength_evo}. Contrary to classical buckling instabilities, the wavelength does not depend on the fiber length (Fig.~\ref{fig:wavelength_evo}~(a)) {which indicates that the instability results from a local mechanism}. However, perturbations are only observable for fibers one order of magnitude larger than $\lambda$ (gray region at low $\ell$ in Fig.~\ref{fig:wavelength_evo}~(a)).\par

The deformation results from the competition between the elastic restoring force ($F_{\rm{elast}}$) and a forcing by the hydrodynamic force ($F_{\rm{visc}}$). By balancing these two forces one can build an elasto-viscous length $\Lambda$ \cite{Gosselin2014, Coq2008, Chakrabarti2020}:

\begin{align}
F_{\rm{elast}}= F_{\rm{visc}} \,\, \Leftrightarrow \,\, EI/\Lambda^2 = \mu u_0 \Lambda \,\,\Leftrightarrow \,\, 
\Lambda = \left(\frac{F_{elast}}{F_{visc}}\right)^{1/3}
\label{eq:elastoviscouslength}
\end{align}
with $u_0$ the flow velocity, $\mu$ the fluid viscosity, $E$ the Young modulus of the fiber and $I = \frac{hw^3}{12}$ its second moment of area. 

To test this scaling we plotted in Fig.~\ref{fig:wavelength_evo}~(b) the evolution of the wavelength as a function of the flow velocity, for two different Young moduli, different channel widths and different fiber lengths. While the wavelength depends on flow velocity and Young modulus, it does not depend on channel width. Increasing $u_0$ or decreasing $E$ results in a decrease of $\lambda$. The dependence of the wavelength on the flow velocity is nicely fitted by a power law $\lambda \propto u_0^{\alpha}$, with $\alpha = -0.2 \pm 0.1$ independent of $E$. These results are in {fair} agreement with the scaling given by equation (\ref{eq:elastoviscouslength}) which predicts a decrease of the wavelength with the flow velocity as $\lambda \propto u_0^{1/3}$ and an increase with the Young modulus. 
Fig.~\ref{fig:wavelength_evo}~(c) gathers all the measurements shown in Fig.~\ref{fig:wavelength_evo}~(b) enriched with experiments where we varied the fiber width keeping the other parameters constant (green points). All wavelengths collapse onto a unique linear curve when plotted as a function of the elasto-viscous length $\Lambda$, confirming $\Lambda$ to be the {relevant} lengthscale of the system. 

We also analyze the phase velocity $V$ and show its proportionality to the mean flow velocity, with channel and fiber width having only a small impact (see Fig.~2 in the supp. data.). \par

\begin{figure*}[ht!]
\centering
\includegraphics[width=1.8 \columnwidth]{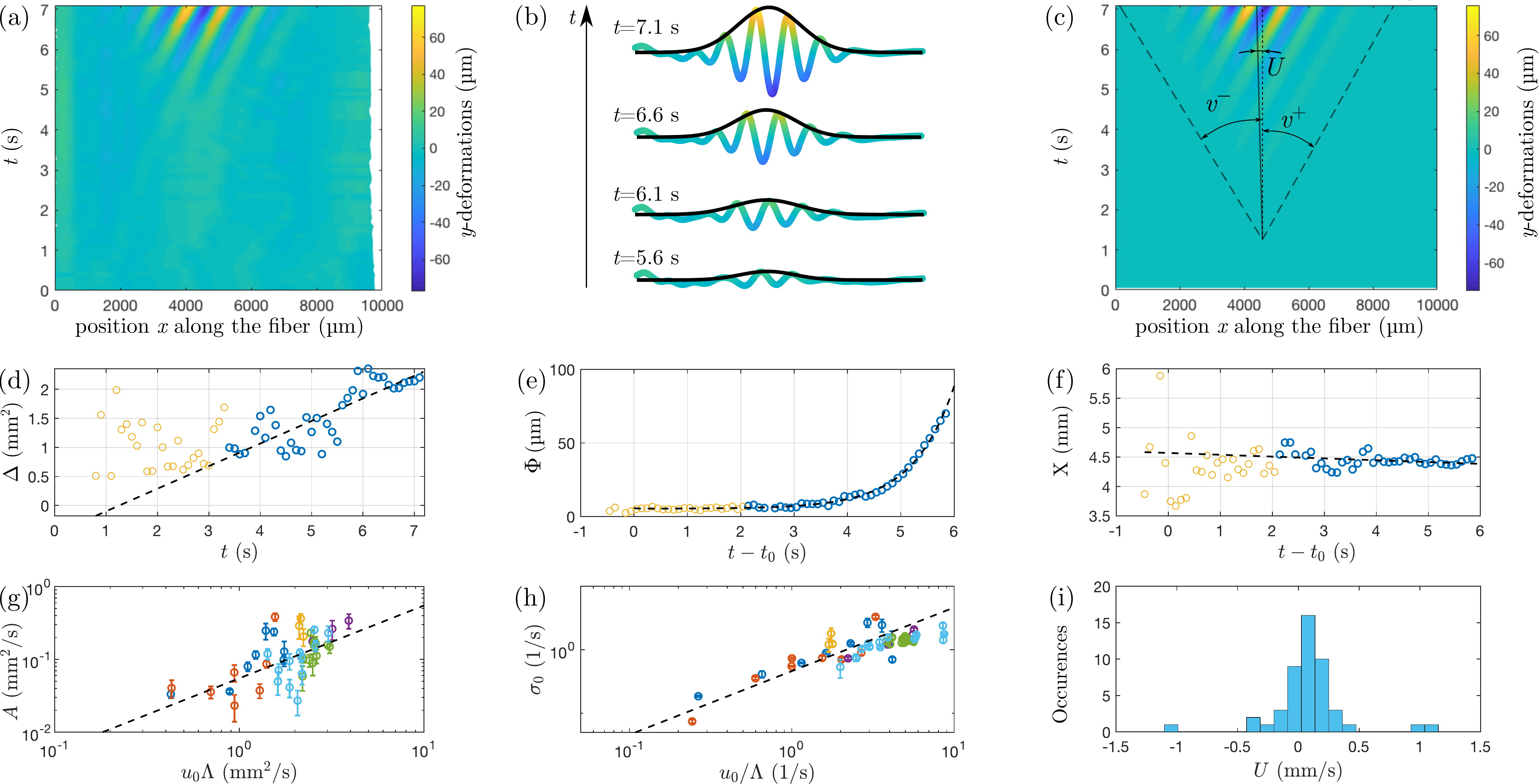}
\caption{(a) Fiber shape as a function of time. Experimental conditions are: $u_0=3.53$ mm/s, $E=27\pm7$ kPa, $w=83\pm3$ $\upmu$m, $h=73\pm3$ $\upmu$m and $\mu=85\pm3$ mPa.s. (b) A few (x, y) shape extracted from (a) illustrating the colour-code used in (a). Black curves are Gaussian fits of the envelope. (c) Same representation as (a) for a Ginzburg-Laudau solution using parameters adjusted from the experiment {shown in (a) and (b)} (see text for details) $A=9.7\pm1.9 \times 10^4$ $\upmu$m$^2$/s, $t_0=1.3\pm0.8$ s, $x_0=4.57\pm0.12$ mm, $U=-31\pm29$ $\upmu$m/s, $\sigma_0=1.37\pm0.07$ s$^{-1}$). $v^+$ and $v^-$ are the velocity of the leading and trailing edges of the wave ($v^{\pm} = U \pm \sqrt{4A\sigma_0}$). (d), (e) and (f) correspond respectively to the evolution of $\Delta$, $\Phi$ and $X$ as a function of time. Dotted black curves represent fits with the equations (\ref{eq:amplitude_G-L2}-\ref{eq:amplitude_G-L4}). Only blue dots are taken into account for the fits, yellow dots are excluded because they derive from measurements of small deformations and are poorly reliable. (g) Diffusion coefficient of the envelope $A$ as a function of $u_0 \Lambda$ and (h) growth rate $\sigma_0$ as a function of $u_0/\Lambda$ fitted from the different experiments shown in Fig.\ref{fig:wavelength_evo} (same color code). The dotted black lines are of slope unity. (i) Histogram of the group velocity $U$. The set of experiments is the same as in Fig.~\ref{fig:wavelength_evo}.}
\label{fig:GL}
\end{figure*} 

To further describe the instability, including the envelop dynamics, we plot the spatio-temporal evolution of the fiber shape (see Fig.~\ref{fig:GL}~(a)). Here, $x$-axis is the position along the fiber and the color code gives the deformation of the fiber in the perpendicular direction while time is represented on the vertical axis from bottom to top. The perturbation is initiated at roughly the center of the fiber - region where potential ends effect are less likely to occur - and grows and spreads as time increases.  
The signal can be described as a traveling wave modulated by an amplitude (see Fig.~\ref{fig:GL}~(b)):
\begin{align}
y(x,t) =\operatorname{Re}\left\{ Y(x,t) e^{i \frac{2 \pi}{\lambda} (x -V t)}\right\},
\label{eq:GL}
\end{align}
with $Y(x,t)$ the amplitude of the envelope, a function of time and position.

On Figs.~\ref{fig:GL}~(a) and (b) we observe that the perturbation eventually invades all the fiber which indicates of an {\it absolute instability}. {Consequently we} use the linear {and real} Ginzburg-Landau (G-L) amplitude equation \cite{Godreche1998} to describe the instability: 
\begin{align}
\frac{\partial Y}{\partial t} = \sigma_0 Y - U \frac{\partial Y}{\partial x}+ A \frac{\partial^2 Y}{\partial x^2}, 
\label{eq:GL}
\end{align}

where $\sigma_0$, $U$ and $A$, are the growth rate, the group velocity, and the diffusion coefficient of the envelope.

The general solution of equation (\ref{eq:GL}), for a Dirac perturbation of amplitude $Y_0$ at time $t_0$ and position $x_0$, is a Gaussian function  \cite{Castaing2005}:

\begin{align}
Y(x,t)=\Phi(t)e^{-\frac{(x-X(t))^2}{\Delta(t)}},
\label{eq:amplitude_G-L}
\end{align}

with the three quantities:
\begin{align}
\Delta(t) &= 4A(t-t_0),
\label{eq:amplitude_G-L2}
\\ \Phi(t) &= -{Y_0}/\left({2 \pi \sqrt{2A(t-t_0)}}\right) e^{\sigma_0(t-t_0)} 
\\
%\label{eq:amplitude_G-L3}
\text{ and } \, \, X(t) &= x_0+ U(t-t_0)
\label{eq:amplitude_G-L4}
\end{align} 
that only depend on time.

 We then fit the envelope extracted from the experiments  by a Gaussian function for every instant (see Fig.~\ref{fig:GL}~(b)) to obtain $\Delta$, $\Phi$ and $X$ as a function of time (see Fig.~\ref{fig:GL}~(d), (e) and (f)). {The temporal evolutions of these parameters are in good agreement with the equations (\ref{eq:amplitude_G-L2}-\ref{eq:amplitude_G-L4}) as shown by the dotted curves in Fig.~\ref{fig:GL}~(d-f) that result from the fitting of the data. The values of  $A$, $t_0$, $\sigma_0$, $U$ and $x_0$ extracted from the fitting procedure are} then used to build the solution of the Ginzburg-Landau equation $Y(x,t)$.  Fig.~\ref{fig:GL}~(c) shows the resulting spatio-temporal evolution which is in excellent agreement  with the experimental observations (Fig.~\ref{fig:GL}~(a)) confirming that this new instability is indeed {\it local and linear. }

To link the ad-hoc description of the system dynamics with the linear G-L equation  to the physics of the underlying problem we determine the evolution of $A$, $\sigma_0$ and $U$ {as a function of} $u_0$, $w$ and $E$.  We observe that the diffusion coefficient $A$ scales reasonably well as $u_0 \Lambda$ (see Fig.~\ref{fig:GL}~(g)). Fig.~\ref{fig:GL}~(h) shows the evolution of the growth rate $\sigma_0$ as a function of $u_0/\Lambda$. All points {collapse fairly well} onto a master curve, confirming that the characteristic length of the system is $\Lambda$ and that all velocities are proportional to $u_0$. Lastly, as visible in Fig.~\ref{fig:GL}~(i), $U$ is always close or equal to zero, i.e. the center of the envelope only slightly moves along the fibers.

\paragraph{Conclusion}

When sufficiently long i.e. $\ell \gg \Lambda$, a parallel fiber transported by an external flow in a confining geometry - the fiber height being similar to the channel height - experiences an instability. The fiber deforms in a wavy shape modulated by an envelope which spreads along the fiber and whose amplitude increases with time. We {show} that the wavelength is constant {with} time and does not depend on the fiber length. It is set instead by the elasto-viscous length $\Lambda$ on which elastic forces balance hydrodynamic forces. By investigating the spatio-temporal evolution of the signal we confirm that we witness an {\it{absolute}} instability arising from a {\it local} amplification mechanism. We show that the dynamics of the amplitude and the spreading of the envelope are well described by the linear Ginzburg-Landau equation and the overall envelope dynamics is completely captured by three parameters: a group velocity $U$, a diffusion coefficient $A$ and a growth rate $\sigma_0$. In all our experiments the group velocity is close to zero, meaning that the envelope only slightly moves along the fibers. On the contrary, the wavelength, the phase velocity, the growth rate and the diffusion coefficient all depend on $\Lambda$ and/or on $u_0$. All velocities are proportional to the flow velocity as expected in viscous flows. As $\Lambda$ takes into account all the physical ingredients of the system we can claim that the instability is fully described by  a competition between two forces: a destabilising hydrodynamic force and a restoring elastic force.
In light of these conclusions we attempted several approaches to derive a dispersion relation (see reference \cite{CappelloThesis}), however further efforts are necessary for a full understanding of the detailed mechanisms.


\begin{thebibliography}{46}
\expandafter\ifx\csname natexlab\endcsname\relax\def\natexlab#1{#1}\fi
\expandafter\ifx\csname bibnamefont\endcsname\relax
  \def\bibnamefont#1{#1}\fi
\expandafter\ifx\csname bibfnamefont\endcsname\relax
  \def\bibfnamefont#1{#1}\fi
\expandafter\ifx\csname citenamefont\endcsname\relax
  \def\citenamefont#1{#1}\fi
\expandafter\ifx\csname url\endcsname\relax
  \def\url#1{\texttt{#1}}\fi
\expandafter\ifx\csname urlprefix\endcsname\relax\def\urlprefix{URL }\fi
\providecommand{\bibinfo}[2]{#2}
\providecommand{\eprint}[2][]{\url{#2}}

\bibitem[{\citenamefont{du~Roure et~al.}(2019)\citenamefont{du~Roure, Lindner,
  Nazockdast, and Shelley}}]{duRoure2019}
\bibinfo{author}{\bibfnamefont{O.}~\bibnamefont{du~Roure}},
  \bibinfo{author}{\bibfnamefont{A.}~\bibnamefont{Lindner}},
  \bibinfo{author}{\bibfnamefont{E.}~\bibnamefont{Nazockdast}},
  \bibnamefont{and} \bibinfo{author}{\bibfnamefont{M.}~\bibnamefont{Shelley}},
  \bibinfo{journal}{Annual Review of Fluid Mechanics}  (\bibinfo{year}{2019}).

\bibitem[{\citenamefont{Cappello et~al.}(2019)\citenamefont{Cappello, Bechert,
  Duprat, du~Roure, Gallaire, and Lindner}}]{Cappello2019}
\bibinfo{author}{\bibfnamefont{J.}~\bibnamefont{Cappello}},
  \bibinfo{author}{\bibfnamefont{M.}~\bibnamefont{Bechert}},
  \bibinfo{author}{\bibfnamefont{C.}~\bibnamefont{Duprat}},
  \bibinfo{author}{\bibfnamefont{O.}~\bibnamefont{du~Roure}},
  \bibinfo{author}{\bibfnamefont{F.}~\bibnamefont{Gallaire}}, \bibnamefont{and}
  \bibinfo{author}{\bibfnamefont{A.}~\bibnamefont{Lindner}},
  \bibinfo{journal}{Physical Review Fluids} \textbf{\bibinfo{volume}{4}},
  \bibinfo{pages}{034202} (\bibinfo{year}{2019}).

\bibitem[{\citenamefont{Marchetti et~al.}(2018)\citenamefont{Marchetti, Raspa,
  Lindner, du~Roure, Bergougnoux, Guazzelli, and Duprat}}]{Marchetti2018}
\bibinfo{author}{\bibfnamefont{B.}~\bibnamefont{Marchetti}},
  \bibinfo{author}{\bibfnamefont{V.}~\bibnamefont{Raspa}},
  \bibinfo{author}{\bibfnamefont{A.}~\bibnamefont{Lindner}},
  \bibinfo{author}{\bibfnamefont{O.}~\bibnamefont{du~Roure}},
  \bibinfo{author}{\bibfnamefont{L.}~\bibnamefont{Bergougnoux}},
  \bibinfo{author}{\bibfnamefont{{\'{E}}.}~\bibnamefont{Guazzelli}},
  \bibnamefont{and} \bibinfo{author}{\bibfnamefont{C.}~\bibnamefont{Duprat}},
  \bibinfo{journal}{to be submitted} pp. \bibinfo{pages}{1--23}
  (\bibinfo{year}{2018}).

\bibitem[{\citenamefont{Li et~al.}(2013)\citenamefont{Li, Manikantan,
  Saintillan, and Spagnolie}}]{Li2013}
\bibinfo{author}{\bibfnamefont{L.}~\bibnamefont{Li}},
  \bibinfo{author}{\bibfnamefont{H.}~\bibnamefont{Manikantan}},
  \bibinfo{author}{\bibfnamefont{D.}~\bibnamefont{Saintillan}},
  \bibnamefont{and} \bibinfo{author}{\bibfnamefont{S.~E.}
  \bibnamefont{Spagnolie}}, \bibinfo{journal}{Journal of Fluid Mechanics}
  \textbf{\bibinfo{volume}{735}}, \bibinfo{pages}{705} (\bibinfo{year}{2013}),
  ISSN \bibinfo{issn}{0022-1120}, \eprint{1306.4692},
  \urlprefix\url{http://www.journals.cambridge.org/abstract{\_}S0022112013005120}.

\bibitem[{\citenamefont{Guglielmini et~al.}(2012)\citenamefont{Guglielmini,
  Kushwaha, Shaqfeh, and Stone}}]{Guglielmini2012}
\bibinfo{author}{\bibfnamefont{L.}~\bibnamefont{Guglielmini}},
  \bibinfo{author}{\bibfnamefont{A.}~\bibnamefont{Kushwaha}},
  \bibinfo{author}{\bibfnamefont{E.~S.} \bibnamefont{Shaqfeh}},
  \bibnamefont{and} \bibinfo{author}{\bibfnamefont{H.~A.} \bibnamefont{Stone}},
  \bibinfo{journal}{Physics of Fluids} \textbf{\bibinfo{volume}{24}},
  \bibinfo{pages}{123601} (\bibinfo{year}{2012}).

\bibitem[{\citenamefont{Young and Shelley}(2007)}]{Young2007}
\bibinfo{author}{\bibfnamefont{Y.-N.} \bibnamefont{Young}} \bibnamefont{and}
  \bibinfo{author}{\bibfnamefont{M.~J.} \bibnamefont{Shelley}},
  \bibinfo{journal}{Physical review letters} \textbf{\bibinfo{volume}{99}},
  \bibinfo{pages}{058303} (\bibinfo{year}{2007}).

\bibitem[{\citenamefont{Wandersman et~al.}(2010)\citenamefont{Wandersman,
  Quennouz, Fermigier, Lindner, and Du~Roure}}]{Wandersman2010}
\bibinfo{author}{\bibfnamefont{E.}~\bibnamefont{Wandersman}},
  \bibinfo{author}{\bibfnamefont{N.}~\bibnamefont{Quennouz}},
  \bibinfo{author}{\bibfnamefont{M.}~\bibnamefont{Fermigier}},
  \bibinfo{author}{\bibfnamefont{A.}~\bibnamefont{Lindner}}, \bibnamefont{and}
  \bibinfo{author}{\bibfnamefont{O.}~\bibnamefont{Du~Roure}},
  \bibinfo{journal}{Soft matter} \textbf{\bibinfo{volume}{6}},
  \bibinfo{pages}{5715} (\bibinfo{year}{2010}).

\bibitem[{\citenamefont{Chakrabarti et~al.}(2020)\citenamefont{Chakrabarti,
  Liu, LaGrone, Cortez, Fauci, du~Roure, Saintillan, and
  Lindner}}]{Chakrabarti2020}
\bibinfo{author}{\bibfnamefont{B.}~\bibnamefont{Chakrabarti}},
  \bibinfo{author}{\bibfnamefont{Y.}~\bibnamefont{Liu}},
  \bibinfo{author}{\bibfnamefont{J.}~\bibnamefont{LaGrone}},
  \bibinfo{author}{\bibfnamefont{R.}~\bibnamefont{Cortez}},
  \bibinfo{author}{\bibfnamefont{L.}~\bibnamefont{Fauci}},
  \bibinfo{author}{\bibfnamefont{O.}~\bibnamefont{du~Roure}},
  \bibinfo{author}{\bibfnamefont{D.}~\bibnamefont{Saintillan}},
  \bibnamefont{and} \bibinfo{author}{\bibfnamefont{A.}~\bibnamefont{Lindner}},
  \bibinfo{journal}{Nature Physics} \textbf{\bibinfo{volume}{16}},
  \bibinfo{pages}{689} (\bibinfo{year}{2020}).

\bibitem[{\citenamefont{Liu et~al.}(2018)\citenamefont{Liu, Chakrabarti,
  Saintillan, Lindner, and Du~Roure}}]{Liu2018}
\bibinfo{author}{\bibfnamefont{Y.}~\bibnamefont{Liu}},
  \bibinfo{author}{\bibfnamefont{B.}~\bibnamefont{Chakrabarti}},
  \bibinfo{author}{\bibfnamefont{D.}~\bibnamefont{Saintillan}},
  \bibinfo{author}{\bibfnamefont{A.}~\bibnamefont{Lindner}}, \bibnamefont{and}
  \bibinfo{author}{\bibfnamefont{O.}~\bibnamefont{Du~Roure}},
  \bibinfo{journal}{Proceedings of the National Academy of Sciences}
  \textbf{\bibinfo{volume}{115}}, \bibinfo{pages}{9438} (\bibinfo{year}{2018}).

\bibitem[{\citenamefont{Forgacs and Mason}(1959)}]{Forgacs1959}
\bibinfo{author}{\bibfnamefont{O.}~\bibnamefont{Forgacs}} \bibnamefont{and}
  \bibinfo{author}{\bibfnamefont{S.}~\bibnamefont{Mason}},
  \bibinfo{journal}{Journal of Colloid Science} \textbf{\bibinfo{volume}{14}},
  \bibinfo{pages}{473} (\bibinfo{year}{1959}).

\bibitem[{\citenamefont{Harasim et~al.}(2013)\citenamefont{Harasim, Wunderlich,
  Peleg, Kr{\"o}ger, and Bausch}}]{Harasim2013}
\bibinfo{author}{\bibfnamefont{M.}~\bibnamefont{Harasim}},
  \bibinfo{author}{\bibfnamefont{B.}~\bibnamefont{Wunderlich}},
  \bibinfo{author}{\bibfnamefont{O.}~\bibnamefont{Peleg}},
  \bibinfo{author}{\bibfnamefont{M.}~\bibnamefont{Kr{\"o}ger}},
  \bibnamefont{and} \bibinfo{author}{\bibfnamefont{A.~R.}
  \bibnamefont{Bausch}}, \bibinfo{journal}{Physical review letters}
  \textbf{\bibinfo{volume}{110}}, \bibinfo{pages}{108302}
  (\bibinfo{year}{2013}).

\bibitem[{\citenamefont{Nguyen and Fauci}(2014)}]{Nguyen2014}
\bibinfo{author}{\bibfnamefont{H.}~\bibnamefont{Nguyen}} \bibnamefont{and}
  \bibinfo{author}{\bibfnamefont{L.}~\bibnamefont{Fauci}},
  \bibinfo{journal}{Journal of The Royal Society Interface}
  \textbf{\bibinfo{volume}{11}}, \bibinfo{pages}{20140314}
  (\bibinfo{year}{2014}).

\bibitem[{\citenamefont{Becker and Shelley}(2001)}]{Becker2001}
\bibinfo{author}{\bibfnamefont{L.~E.} \bibnamefont{Becker}} \bibnamefont{and}
  \bibinfo{author}{\bibfnamefont{M.~J.} \bibnamefont{Shelley}},
  \bibinfo{journal}{Physical Review Letters} \textbf{\bibinfo{volume}{87}},
  \bibinfo{pages}{198301} (\bibinfo{year}{2001}).

\bibitem[{\citenamefont{Quennouz et~al.}(2015)\citenamefont{Quennouz, Shelley,
  Du~Roure, and Lindner}}]{Quennouz2015}
\bibinfo{author}{\bibfnamefont{N.}~\bibnamefont{Quennouz}},
  \bibinfo{author}{\bibfnamefont{M.}~\bibnamefont{Shelley}},
  \bibinfo{author}{\bibfnamefont{O.}~\bibnamefont{Du~Roure}}, \bibnamefont{and}
  \bibinfo{author}{\bibfnamefont{A.}~\bibnamefont{Lindner}},
  \bibinfo{journal}{J Fluid Mech} \textbf{\bibinfo{volume}{769}},
  \bibinfo{pages}{387} (\bibinfo{year}{2015}).

\bibitem[{\citenamefont{Manikantan and Saintillan}(2013)}]{Manikantan2013}
\bibinfo{author}{\bibfnamefont{H.}~\bibnamefont{Manikantan}} \bibnamefont{and}
  \bibinfo{author}{\bibfnamefont{D.}~\bibnamefont{Saintillan}},
  \bibinfo{journal}{Physics of Fluids} \textbf{\bibinfo{volume}{25}},
  \bibinfo{pages}{073603} (\bibinfo{year}{2013}).

\bibitem[{\citenamefont{Steinhauser et~al.}(2012)\citenamefont{Steinhauser,
  Köster, and Pfohl}}]{Steinhauser2012}
\bibinfo{author}{\bibfnamefont{D.}~\bibnamefont{Steinhauser}},
  \bibinfo{author}{\bibfnamefont{S.}~\bibnamefont{Köster}}, \bibnamefont{and}
  \bibinfo{author}{\bibfnamefont{T.}~\bibnamefont{Pfohl}},
  \bibinfo{journal}{ACS Macro Letters} \textbf{\bibinfo{volume}{1}},
  \bibinfo{pages}{541} (\bibinfo{year}{2012}).

\bibitem[{\citenamefont{Manikantan and Saintillan}(2015)}]{Manikantan2015}
\bibinfo{author}{\bibfnamefont{H.}~\bibnamefont{Manikantan}} \bibnamefont{and}
  \bibinfo{author}{\bibfnamefont{D.}~\bibnamefont{Saintillan}},
  \bibinfo{journal}{Physical Review E} \textbf{\bibinfo{volume}{92}},
  \bibinfo{pages}{041002} (\bibinfo{year}{2015}).

\bibitem[{\citenamefont{Liu et~al.}(2020)\citenamefont{Liu, Zografos, Fidalgo,
  Duch{\^e}ne, Quintard, Darnige, Vasco, Huille, du~Roure, Oliveira
  et~al.}}]{Liu2020}
\bibinfo{author}{\bibfnamefont{Y.}~\bibnamefont{Liu}},
  \bibinfo{author}{\bibfnamefont{K.}~\bibnamefont{Zografos}},
  \bibinfo{author}{\bibfnamefont{J.}~\bibnamefont{Fidalgo}},
  \bibinfo{author}{\bibfnamefont{C.}~\bibnamefont{Duch{\^e}ne}},
  \bibinfo{author}{\bibfnamefont{C.}~\bibnamefont{Quintard}},
  \bibinfo{author}{\bibfnamefont{T.}~\bibnamefont{Darnige}},
  \bibinfo{author}{\bibfnamefont{F.}~\bibnamefont{Vasco}},
  \bibinfo{author}{\bibfnamefont{S.}~\bibnamefont{Huille}},
  \bibinfo{author}{\bibfnamefont{O.}~\bibnamefont{du~Roure}},
  \bibinfo{author}{\bibfnamefont{M.~S.} \bibnamefont{Oliveira}},
  \bibnamefont{et~al.}, \bibinfo{journal}{Soft Matter}  (\bibinfo{year}{2020}).

\bibitem[{\citenamefont{Xu and Nadim}(1994)}]{Xu1994}
\bibinfo{author}{\bibfnamefont{X.}~\bibnamefont{Xu}} \bibnamefont{and}
  \bibinfo{author}{\bibfnamefont{A.}~\bibnamefont{Nadim}},
  \bibinfo{journal}{Physics of Fluids} \textbf{\bibinfo{volume}{6}},
  \bibinfo{pages}{2889} (\bibinfo{year}{1994}), ISSN \bibinfo{issn}{1070-6631},
  \urlprefix\url{http://scitation.aip.org/content/aip/journal/pof2/6/9/10.1063/1.868116}.

\bibitem[{\citenamefont{D'Angelo
  et~al.}(2009{\natexlab{a}})\citenamefont{D'Angelo, Semin, Picard, Poitzsch,
  Hulin, and Auradou}}]{DAngelo2009_1}
\bibinfo{author}{\bibfnamefont{M.~V.} \bibnamefont{D'Angelo}},
  \bibinfo{author}{\bibfnamefont{B.}~\bibnamefont{Semin}},
  \bibinfo{author}{\bibfnamefont{G.}~\bibnamefont{Picard}},
  \bibinfo{author}{\bibfnamefont{M.~E.} \bibnamefont{Poitzsch}},
  \bibinfo{author}{\bibfnamefont{J.~P.} \bibnamefont{Hulin}}, \bibnamefont{and}
  \bibinfo{author}{\bibfnamefont{H.}~\bibnamefont{Auradou}},
  \bibinfo{journal}{Trans. Porous Med.} \textbf{\bibinfo{volume}{84}},
  \bibinfo{pages}{389} (\bibinfo{year}{2009}{\natexlab{a}}), ISSN
  \bibinfo{issn}{0169-3913},
  \urlprefix\url{http://link.springer.com/10.1007/s11242-009-9507-x}.

\bibitem[{\citenamefont{D'Angelo
  et~al.}(2009{\natexlab{b}})\citenamefont{D'Angelo, Auradou, Picard, Poitzsch,
  and Hulin}}]{dAngelo2009}
\bibinfo{author}{\bibfnamefont{M.}~\bibnamefont{D'Angelo}},
  \bibinfo{author}{\bibfnamefont{H.}~\bibnamefont{Auradou}},
  \bibinfo{author}{\bibfnamefont{G.}~\bibnamefont{Picard}},
  \bibinfo{author}{\bibfnamefont{M.}~\bibnamefont{Poitzsch}}, \bibnamefont{and}
  \bibinfo{author}{\bibfnamefont{J.}~\bibnamefont{Hulin}}, in
  \emph{\bibinfo{booktitle}{Journal of Physics: Conference Series}}
  (\bibinfo{organization}{IOP Publishing}, \bibinfo{year}{2009}{\natexlab{b}}),
  vol. \bibinfo{volume}{166}, p. \bibinfo{pages}{012001}.

\bibitem[{\citenamefont{Triantafyllou et~al.}(1987)\citenamefont{Triantafyllou,
  Kupfer, and Bers}}]{Triantafyllou1987}
\bibinfo{author}{\bibfnamefont{G.~S.} \bibnamefont{Triantafyllou}},
  \bibinfo{author}{\bibfnamefont{K.}~\bibnamefont{Kupfer}}, \bibnamefont{and}
  \bibinfo{author}{\bibfnamefont{A.}~\bibnamefont{Bers}},
  \bibinfo{journal}{Physical review letters} \textbf{\bibinfo{volume}{59}},
  \bibinfo{pages}{1914} (\bibinfo{year}{1987}).

\bibitem[{\citenamefont{Gondret et~al.}(1999)\citenamefont{Gondret, Ern,
  Meignin, and Rabaud}}]{Gondret1999}
\bibinfo{author}{\bibfnamefont{P.}~\bibnamefont{Gondret}},
  \bibinfo{author}{\bibfnamefont{P.}~\bibnamefont{Ern}},
  \bibinfo{author}{\bibfnamefont{L.}~\bibnamefont{Meignin}}, \bibnamefont{and}
  \bibinfo{author}{\bibfnamefont{M.}~\bibnamefont{Rabaud}},
  \bibinfo{journal}{Physical review letters} \textbf{\bibinfo{volume}{82}},
  \bibinfo{pages}{1442} (\bibinfo{year}{1999}).

\bibitem[{\citenamefont{Charru}(2012)}]{Charru2012}
\bibinfo{author}{\bibfnamefont{F.}~\bibnamefont{Charru}},
  \emph{\bibinfo{title}{Instabilit{\'e}s hydrodynamiques}}
  (\bibinfo{publisher}{EDP Sciences}, \bibinfo{year}{2012}).

\bibitem[{\citenamefont{Huerre et~al.}(2000)\citenamefont{Huerre, Batchelor,
  Moffatt, and Worster}}]{Huerre2000}
\bibinfo{author}{\bibfnamefont{P.}~\bibnamefont{Huerre}},
  \bibinfo{author}{\bibfnamefont{G.}~\bibnamefont{Batchelor}},
  \bibinfo{author}{\bibfnamefont{H.}~\bibnamefont{Moffatt}}, \bibnamefont{and}
  \bibinfo{author}{\bibfnamefont{M.}~\bibnamefont{Worster}},
  \bibinfo{journal}{Perspectives in fluid dynamics} pp.
  \bibinfo{pages}{159--229} (\bibinfo{year}{2000}).

\bibitem[{\citenamefont{Utada et~al.}(2008)\citenamefont{Utada,
  Fernandez-Nieves, Gordillo, and Weitz}}]{Utada2008}
\bibinfo{author}{\bibfnamefont{A.~S.} \bibnamefont{Utada}},
  \bibinfo{author}{\bibfnamefont{A.}~\bibnamefont{Fernandez-Nieves}},
  \bibinfo{author}{\bibfnamefont{J.~M.} \bibnamefont{Gordillo}},
  \bibnamefont{and} \bibinfo{author}{\bibfnamefont{D.~A.} \bibnamefont{Weitz}},
  \bibinfo{journal}{Physical review letters} \textbf{\bibinfo{volume}{100}},
  \bibinfo{pages}{014502} (\bibinfo{year}{2008}).

\bibitem[{\citenamefont{Gordillo et~al.}(2001)\citenamefont{Gordillo,
  Ga{\~n}{\'a}n-Calvo, and P{\'e}rez-Saborid}}]{Gordillo2001}
\bibinfo{author}{\bibfnamefont{J.~M.} \bibnamefont{Gordillo}},
  \bibinfo{author}{\bibfnamefont{A.~M.} \bibnamefont{Ga{\~n}{\'a}n-Calvo}},
  \bibnamefont{and}
  \bibinfo{author}{\bibfnamefont{M.}~\bibnamefont{P{\'e}rez-Saborid}},
  \bibinfo{journal}{Physics of fluids} \textbf{\bibinfo{volume}{13}},
  \bibinfo{pages}{3839} (\bibinfo{year}{2001}).

\bibitem[{\citenamefont{Ga{\~n}{\'a}n-Calvo et~al.}(2006)}]{Ganan2006}
\bibinfo{author}{\bibfnamefont{A.~M.} \bibnamefont{Ga{\~n}{\'a}n-Calvo}}
  \bibnamefont{et~al.}, \bibinfo{journal}{Journal of Fluid Mechanics}
  \textbf{\bibinfo{volume}{553}}, \bibinfo{pages}{75} (\bibinfo{year}{2006}).

\bibitem[{\citenamefont{Guillot et~al.}(2007)\citenamefont{Guillot, Colin,
  Utada, and Ajdari}}]{Guillot2007}
\bibinfo{author}{\bibfnamefont{P.}~\bibnamefont{Guillot}},
  \bibinfo{author}{\bibfnamefont{A.}~\bibnamefont{Colin}},
  \bibinfo{author}{\bibfnamefont{A.~S.} \bibnamefont{Utada}}, \bibnamefont{and}
  \bibinfo{author}{\bibfnamefont{A.}~\bibnamefont{Ajdari}},
  \bibinfo{journal}{Physical review letters} \textbf{\bibinfo{volume}{99}},
  \bibinfo{pages}{104502} (\bibinfo{year}{2007}).

\bibitem[{\citenamefont{Duprat et~al.}(2007)\citenamefont{Duprat, Ruyer-Quil,
  Kalliadasis, and Giorgiutti-Dauphin{\'e}}}]{Duprat2007}
\bibinfo{author}{\bibfnamefont{C.}~\bibnamefont{Duprat}},
  \bibinfo{author}{\bibfnamefont{C.}~\bibnamefont{Ruyer-Quil}},
  \bibinfo{author}{\bibfnamefont{S.}~\bibnamefont{Kalliadasis}},
  \bibnamefont{and}
  \bibinfo{author}{\bibfnamefont{F.}~\bibnamefont{Giorgiutti-Dauphin{\'e}}},
  \bibinfo{journal}{Physical review letters} \textbf{\bibinfo{volume}{98}},
  \bibinfo{pages}{244502} (\bibinfo{year}{2007}).

\bibitem[{\citenamefont{Gallaire and Brun}(2017)}]{Gallaire2017}
\bibinfo{author}{\bibfnamefont{F.}~\bibnamefont{Gallaire}} \bibnamefont{and}
  \bibinfo{author}{\bibfnamefont{P.-T.} \bibnamefont{Brun}},
  \bibinfo{journal}{Philosophical Transactions of the Royal Society A:
  Mathematical, Physical and Engineering Sciences}
  \textbf{\bibinfo{volume}{375}}, \bibinfo{pages}{20160155}
  (\bibinfo{year}{2017}).

\bibitem[{\citenamefont{Huerre and Monkewitz}(1990)}]{Huerre1990}
\bibinfo{author}{\bibfnamefont{P.}~\bibnamefont{Huerre}} \bibnamefont{and}
  \bibinfo{author}{\bibfnamefont{P.~A.} \bibnamefont{Monkewitz}},
  \bibinfo{journal}{Annual review of fluid mechanics}
  \textbf{\bibinfo{volume}{22}}, \bibinfo{pages}{473} (\bibinfo{year}{1990}).

\bibitem[{\citenamefont{Dendukuri et~al.}(2008)\citenamefont{Dendukuri, Panda,
  Haghgooie, Kim, Hatton, and Doyle}}]{Dendukuri2008}
\bibinfo{author}{\bibfnamefont{D.}~\bibnamefont{Dendukuri}},
  \bibinfo{author}{\bibfnamefont{P.}~\bibnamefont{Panda}},
  \bibinfo{author}{\bibfnamefont{R.}~\bibnamefont{Haghgooie}},
  \bibinfo{author}{\bibfnamefont{J.~M.} \bibnamefont{Kim}},
  \bibinfo{author}{\bibfnamefont{T.~A.} \bibnamefont{Hatton}},
  \bibnamefont{and} \bibinfo{author}{\bibfnamefont{P.~S.} \bibnamefont{Doyle}},
  \bibinfo{journal}{Macromolecules} \textbf{\bibinfo{volume}{41}},
  \bibinfo{pages}{8547} (\bibinfo{year}{2008}), ISSN \bibinfo{issn}{00249297}.

\bibitem[{\citenamefont{Dendukuri et~al.}(2007)\citenamefont{Dendukuri, Gu,
  Pregibon, Hatton, and Doyle}}]{Dendukuri2007}
\bibinfo{author}{\bibfnamefont{D.}~\bibnamefont{Dendukuri}},
  \bibinfo{author}{\bibfnamefont{S.~S.} \bibnamefont{Gu}},
  \bibinfo{author}{\bibfnamefont{D.~C.} \bibnamefont{Pregibon}},
  \bibinfo{author}{\bibfnamefont{T.~A.} \bibnamefont{Hatton}},
  \bibnamefont{and} \bibinfo{author}{\bibfnamefont{P.~S.} \bibnamefont{Doyle}},
  \bibinfo{journal}{Lab on a Chip} \textbf{\bibinfo{volume}{7}},
  \bibinfo{pages}{818} (\bibinfo{year}{2007}), ISSN \bibinfo{issn}{1473-0197},
  \urlprefix\url{http://xlink.rsc.org/?DOI=b703457a}.

\bibitem[{\citenamefont{Duprat et~al.}(2014)\citenamefont{Duprat, Berthet,
  Wexler, du~Roure, and Lindner}}]{Duprat2014}
\bibinfo{author}{\bibfnamefont{C.}~\bibnamefont{Duprat}},
  \bibinfo{author}{\bibfnamefont{H.}~\bibnamefont{Berthet}},
  \bibinfo{author}{\bibfnamefont{J.~S.} \bibnamefont{Wexler}},
  \bibinfo{author}{\bibfnamefont{O.}~\bibnamefont{du~Roure}}, \bibnamefont{and}
  \bibinfo{author}{\bibfnamefont{A.}~\bibnamefont{Lindner}},
  \bibinfo{journal}{Lab on a chip} \textbf{\bibinfo{volume}{15}},
  \bibinfo{pages}{244} (\bibinfo{year}{2014}), ISSN \bibinfo{issn}{1473-0189},
  \urlprefix\url{http://www.ncbi.nlm.nih.gov/pubmed/25360871}.

\bibitem[{\citenamefont{Cappello et~al.}(2020)\citenamefont{Cappello,
  D’herbemont, Lindner, and Du~Roure}}]{Cappello2020}
\bibinfo{author}{\bibfnamefont{J.}~\bibnamefont{Cappello}},
  \bibinfo{author}{\bibfnamefont{V.}~\bibnamefont{D’herbemont}},
  \bibinfo{author}{\bibfnamefont{A.}~\bibnamefont{Lindner}}, \bibnamefont{and}
  \bibinfo{author}{\bibfnamefont{O.}~\bibnamefont{Du~Roure}},
  \bibinfo{journal}{Micromachines} \textbf{\bibinfo{volume}{11}},
  \bibinfo{pages}{318} (\bibinfo{year}{2020}).

\bibitem[{\citenamefont{Wexler et~al.}(2013)\citenamefont{Wexler, Trinh,
  Berthet, Quennouz, du~Roure, Huppert, Linder, and Stone}}]{Wexler2013}
\bibinfo{author}{\bibfnamefont{J.~S.} \bibnamefont{Wexler}},
  \bibinfo{author}{\bibfnamefont{P.~H.} \bibnamefont{Trinh}},
  \bibinfo{author}{\bibfnamefont{H.}~\bibnamefont{Berthet}},
  \bibinfo{author}{\bibfnamefont{N.}~\bibnamefont{Quennouz}},
  \bibinfo{author}{\bibfnamefont{O.}~\bibnamefont{du~Roure}},
  \bibinfo{author}{\bibfnamefont{H.~E.} \bibnamefont{Huppert}},
  \bibinfo{author}{\bibfnamefont{A.}~\bibnamefont{Linder}}, \bibnamefont{and}
  \bibinfo{author}{\bibfnamefont{H.~A.} \bibnamefont{Stone}},
  \bibinfo{journal}{Journal of Fluid Mechanics} \textbf{\bibinfo{volume}{720}},
  \bibinfo{pages}{517} (\bibinfo{year}{2013}), ISSN \bibinfo{issn}{0022-1120},
  \urlprefix\url{http://www.journals.cambridge.org/abstract{\_}S0022112013000499}.

\bibitem[{\citenamefont{Berthet et~al.}(2016)\citenamefont{Berthet, du~Roure,
  and Lindner}}]{Berthet2016}
\bibinfo{author}{\bibfnamefont{H.}~\bibnamefont{Berthet}},
  \bibinfo{author}{\bibfnamefont{O.}~\bibnamefont{du~Roure}}, \bibnamefont{and}
  \bibinfo{author}{\bibfnamefont{A.}~\bibnamefont{Lindner}},
  \bibinfo{journal}{Applied Sciences} \textbf{\bibinfo{volume}{6}},
  \bibinfo{pages}{385} (\bibinfo{year}{2016}), ISSN \bibinfo{issn}{2076-3417},
  \urlprefix\url{http://www.mdpi.com/2076-3417/6/12/385}.

\bibitem[{\citenamefont{Nagel et~al.}(2018)\citenamefont{Nagel, Brun, Berthet,
  Lindner, Gallaire, and Duprat}}]{Nagel2018}
\bibinfo{author}{\bibfnamefont{M.}~\bibnamefont{Nagel}},
  \bibinfo{author}{\bibfnamefont{P.-T.} \bibnamefont{Brun}},
  \bibinfo{author}{\bibfnamefont{H.}~\bibnamefont{Berthet}},
  \bibinfo{author}{\bibfnamefont{A.}~\bibnamefont{Lindner}},
  \bibinfo{author}{\bibfnamefont{F.}~\bibnamefont{Gallaire}}, \bibnamefont{and}
  \bibinfo{author}{\bibfnamefont{C.}~\bibnamefont{Duprat}},
  \bibinfo{journal}{Journal of Fluid Mechanics} \textbf{\bibinfo{volume}{835}},
  \bibinfo{pages}{444} (\bibinfo{year}{2018}).

\bibitem[{\citenamefont{Bechert et~al.}(2019)\citenamefont{Bechert, Cappello,
  Da{\"\i}eff, Gallaire, Lindner, and Duprat}}]{Bechert2019}
\bibinfo{author}{\bibfnamefont{M.}~\bibnamefont{Bechert}},
  \bibinfo{author}{\bibfnamefont{J.}~\bibnamefont{Cappello}},
  \bibinfo{author}{\bibfnamefont{M.}~\bibnamefont{Da{\"\i}eff}},
  \bibinfo{author}{\bibfnamefont{F.}~\bibnamefont{Gallaire}},
  \bibinfo{author}{\bibfnamefont{A.}~\bibnamefont{Lindner}}, \bibnamefont{and}
  \bibinfo{author}{\bibfnamefont{C.}~\bibnamefont{Duprat}},
  \bibinfo{journal}{EPL (Europhysics Letters)} \textbf{\bibinfo{volume}{126}},
  \bibinfo{pages}{44001} (\bibinfo{year}{2019}).

\bibitem[{\citenamefont{Schneider et~al.}(2012)\citenamefont{Schneider,
  Rasband, and Eliceiri}}]{Schneider2012}
\bibinfo{author}{\bibfnamefont{C.~A.} \bibnamefont{Schneider}},
  \bibinfo{author}{\bibfnamefont{W.~S.} \bibnamefont{Rasband}},
  \bibnamefont{and} \bibinfo{author}{\bibfnamefont{K.~W.}
  \bibnamefont{Eliceiri}}, \bibinfo{journal}{Nature Methods}
  \textbf{\bibinfo{volume}{9}}, \bibinfo{pages}{671} (\bibinfo{year}{2012}),
  ISSN \bibinfo{issn}{1548-7091},
  \urlprefix\url{http://dx.doi.org/10.1038/nmeth.2089}.

\bibitem[{\citenamefont{Gosselin et~al.}(2014)\citenamefont{Gosselin, Neetzow,
  and Paak}}]{Gosselin2014}
\bibinfo{author}{\bibfnamefont{F.}~\bibnamefont{Gosselin}},
  \bibinfo{author}{\bibfnamefont{P.}~\bibnamefont{Neetzow}}, \bibnamefont{and}
  \bibinfo{author}{\bibfnamefont{M.}~\bibnamefont{Paak}},
  \bibinfo{journal}{Physical Review E} \textbf{\bibinfo{volume}{90}},
  \bibinfo{pages}{052718} (\bibinfo{year}{2014}).

\bibitem[{\citenamefont{Coq et~al.}(2008)\citenamefont{Coq, Du~Roure,
  Marthelot, Bartolo, and Fermigier}}]{Coq2008}
\bibinfo{author}{\bibfnamefont{N.}~\bibnamefont{Coq}},
  \bibinfo{author}{\bibfnamefont{O.}~\bibnamefont{Du~Roure}},
  \bibinfo{author}{\bibfnamefont{J.}~\bibnamefont{Marthelot}},
  \bibinfo{author}{\bibfnamefont{D.}~\bibnamefont{Bartolo}}, \bibnamefont{and}
  \bibinfo{author}{\bibfnamefont{M.}~\bibnamefont{Fermigier}},
  \bibinfo{journal}{Physics of Fluids} \textbf{\bibinfo{volume}{20}},
  \bibinfo{pages}{051703} (\bibinfo{year}{2008}).

\bibitem[{\citenamefont{Godreche and Manneville}(1998)}]{Godreche1998}
\bibinfo{author}{\bibfnamefont{C.}~\bibnamefont{Godreche}} \bibnamefont{and}
  \bibinfo{author}{\bibfnamefont{P.}~\bibnamefont{Manneville}},
  \bibinfo{journal}{Oceanographic Literature Review}
  \textbf{\bibinfo{volume}{9}}, \bibinfo{pages}{1723} (\bibinfo{year}{1998}).

\bibitem[{\citenamefont{Castaing}(2005)}]{Castaing2005}
\bibinfo{author}{\bibfnamefont{B.}~\bibnamefont{Castaing}},
  \emph{\bibinfo{title}{Hydrodynamics and nonlinear instabilities}},
  vol.~\bibinfo{volume}{3} (\bibinfo{publisher}{Cambridge University Press},
  \bibinfo{year}{2005}).

\bibitem[{\citenamefont{Cappello}(2020)}]{CappelloThesis}
\bibinfo{author}{\bibfnamefont{J.}~\bibnamefont{Cappello}}, \bibinfo{type}{Phd
  thesis}, \bibinfo{school}{{Universit{\'e} de Paris}} (\bibinfo{year}{2020}),
  \urlprefix\url{https://tel.archives-ouvertes.fr/tel-03260710}.

\end{thebibliography}
\end{document}